%% file: main.tex
\documentclass[conference]{sig-alternate-10pt}

\usepackage{amsmath}
\usepackage[utf8x]{inputenc}
\usepackage{url}
\usepackage{flushend}
\usepackage{graphicx}
\usepackage{booktabs} 

\begin{document}
\title{Spatial Interference Detection for Mobile Visible Light Communication}

\author{Ali Ugur Guler, Tristan Braud and Pan Hui\vspace{0.3em}\\
\large{\{auguler@ust.hk, braudt@ust.hk, panhui@cse.ust.hk\}}\vspace{0.5em}
\\{The Hong Kong University of Science and Technology}}

\maketitle

\begin{abstract}
Taking advantage of the rolling shutter effect of CMOS cameras in smartphones is a common practice
to increase the transfered data rate with visible light communication (VLC) without employing
external equipment such as photodiodes. 
VLC can then be used as replacement of other marker based techniques for object identification for
Augmented Reality and Ubiquitous computing applications. 
However, the rolling shutter effect only allows to transmit data over a single dimension, which
considerably limits the available bandwidth. 
In this article we propose a new method exploiting spacial interference detection to enable parallel
transmission and design a protocol that enables easy identification of interferences between
two signals.
By introducing a second dimension, we are not only able to significantly increase the available
bandwidth, but also identify and isolate light sources in close proximity.
\end{abstract}


\input{introduction}
\input{related_works}

\input{system_design}

\input{evaluation}

\input{conclusion}

\bibliographystyle{abbrv}
\bibliography{main} 

\end{document}

%% file: introduction.tex
\section{Introduction}

The recent emergence of new technologies such as Augmented Reality (AR) or the Internet of Things
(IoT) introduced new use cases which heavily rely on unique object identification. 
For instance, an
application may need to spatially identify several connected devices in order display informations
and controls to the user.
In such a scenario, multiple objects have to be recognized in real time using the camera of a
smartphone or smartglasses, in order to superimpose a virtual control layer on top of the physical
world.

Two main approaches are currently used for object identification: markers that convey a message to the receiver and markerless systems that use image processing techniques~\cite{Genc:2002:MTA:850976.854956}.
Markerless systems rely on computation-heavy vision algorithm to identify devices. Their complexity
dramatically increase when several similar looking devices have to be uniquely identified. 
Using markers therefore presents various benefits. If most ongoing techniques rely on image
recognition, the detection algorithms are lightweight enough to operate seamlessly on mobile
devices. 
Besides, each marker transmits a specific message. This latter property is especially useful for
building ubiquitous networks and their related AR applications, as those networks may integrate
similar looking devices, or even non visible appliances embedded inside the walls.

QR codes are the most widely deployed marker based approach to identify such objects due to their
ability to contain large amounts of information combined with deployment simplicity. 
However, QR codes present several critical drawbacks. 
First of all, QR codes is not aesthetically pleasing. 
For instance, in a home IoT situation, users may not want to cover their walls with markers for the sake of efficiency~\cite{wein2014visual}. 
Second, depending on the position of the marker, it may be tricky for a human to figure out which object it's related to, as the format doesn't present any additional human readable information. 
Finally, as the technology is vision-based, the markers require an unobstructed line of sight, with a small enough viewing angle to be decoded. Similarly, transmitting over longer distances requires QR codes of
increasing sizes.

Visible Light Communication(VLC) may then be considered to solve such issues.
By modulating the intensity of a light source at a higher frequency than the human eye can detect, VLC can be inconspicuously introduced in various environments, for instance in the lighting or
status LEDs of the corresponding object.
Moreover, distance, viewing angle and line of sight will cause VLC to operate with degraded performance to some extent, as opposed as the \textit{all-or-nothing} behavior of QR codes.

In this article, we propose to increase the effectiveness of current VLC based object
identification through parallel transmission. If this solution displays several obvious advantages (easy identification of separate objects, increased bandwidth), it also introduces new challenges to address:
\begin{itemize}
\item identifying which light sources are linked together.
\item isolating the regions of different transmission. 
\item identifying objects close to each other by recognition of the interference region.
\end{itemize}
To face those challenges, we  design a specific data-link layer protocol. By employing orthogonal preambles, we are able to identify the interference regions and isolate the different signals.

This article is organized as follows: after reviewing research studies related to VLC and marker-based object identification (section~\ref{sec:related}), we describe our system design in section~\ref{sec:system}. We finally analyze the performances and limitations of our model in section~\ref{sec:eval}.

%% file: related_works.tex
\section{Related Works}
\label{sec:related}

Marker systems based on VLC can be divided into three categories: color codes, spacial codes and temporal codes.

Color codes are very vulnerable to lighting variations. While they can be used alone for tracking
algorithms, they only provide very small data rates~\cite{koutaki2015poster}. 
For this reason they are generally used in combination with temporal or spatial codes to provide additional information~\cite{Fath:14,Hu:2015:CID:2716281.2836097}.

One of the most used spatial code is the QR code. QR codes have the benefit of high data density and easy application\cite{ISO18004}. However they are prone to errors caused by lighting and orientation changes. Another study proposed to exploit light reflection only to identify moving vehicles~\cite{Wang:2016:PCA:2999572.2999584}.
    
Temporal codes provide more robustness against lighting and orientation variation. 
By modulating LEDs with very high frequency, several teams managed to send data at rates reaching gigabit/s~\cite{azhar2013gigabit,tsonev20143}. 
However, such high data rates can only be achieved with silicon photo diodes. 
On the other hand, the camera on a smartphone can only provide a connection with low data rates~\cite{jovicic2013visible}. 
Even though communication with a CMOS camera is much slower, there are several reasons to prefer using it over photo diodes for object identification:
\begin{itemize}
\item Photo diodes lack the spatial information that cameras have.
\item CMOS cameras are already built in smartphones, while fast response photo diodes are not readily available.
\end{itemize}
Due to Nyquist theorem~\cite{oppenheim1999discrete}, temporal communication through cameras over frames can only achieve a data rate of 15\,Bit/s if the camera can record 30 frames per second. 
Nevertheless, Christos Danakis et al. have showed that by exploiting the rolling shutter effect on CMOS cameras, it is possible to increase the data rate to 1\,Kb/s~\cite{danakis2012using}.

The rolling shutter effect has opened many possibilities for VLC applications on smartphones. Panasonic already has a commercial VLC marker project "Light ID"~\cite{lightid}. 
Luxapose uses the rolling shutter effect to identify different LEDs and use them for accurate indoor positioning~\cite{kuo2014luxapose}. 
Rajagopal et al. have used modulated ambient light in the 8Khz frequency range (making the flickering invisible to the human eye), and successfully separated 29 channels in the time domain with 0.2khz channel separation using rolling shutter effect~\cite{rajagopal2014visual}.
DisCo group has constructed a system that introduces robustness against occlusion, movement and distance~\cite{jo2014disco}.

Several studies have been performed on combining spatial and time domain to increase visible light communication transmission rates.
VRCodes uses binary coding with color, time and space domain to increase data rates by modulating LCD screens or projectors~\cite{woo2012vrcodes}. 
COBRA uses a color barcode stream system to utilize color, time, space domain~\cite{hao2012cobra}.
T. Langlotz and O. Bimber also proposed 4D barcodes projected by screens~\cite{langlotz2007unsynchronized}.
However these works focus on transmission through lcd screens. 

In our work we introduce a method using orthogonal preamble to separate channels on spatial domain.
Our work focuses on identifying interference regions of multiple LEDs used for illumination and introduce spatial parallelism to VLC using LEDs.

%% file: system_design.tex
\section{System Design}
\label{sec:system}

In this study, we consider a scenario for AR applications interacting with ubiquitous computing networks. 
In this arrangement, a large amount devices have to be identified in real time. The conditions of detection may vary from optimal -- marker at close distance, right in front of the camera -- to the marker being partially occulted.
In order to achieve those goals, we exploit the rolling shutter effect of the smartphone's camera, in combination with a specific protocol for the message to be recovered by the receiver.

\subsection{The rolling shutter effect}

Most CMOS cameras integrated in smartphones take pictures using a rolling shutter. The scene is therefore scanned in one direction, line by line, instead of taking a snapshot of the whole scene. 
As all parts of the scene are not recorded at the same instant, fast moving objects and rapid variation of light leave a distinctive, predictable pattern on the image.

By flashing the LED at a high frequency (around 10\,KHz) on the transmitter side, and exploiting
the rolling shutter effect on the receiver side, the light source will leave specific black stripes on the picture, each stripe corresponding to a moment the LED is switched off.

Exploiting this effet, we are able to squeeze more data bits in a single frame of the camera (in our protocol, 28), increasing the effective bandwidth of the system.
Another advantage of this technique resides in the fact that the camera doesn't have to be directly pointed towards the light source. In our case, we use the reflected light on a surface as the main transmission medium.

\subsection{Transmitter}

\begin{figure}[ht]
\centering
\includegraphics[width=0.4\textwidth]{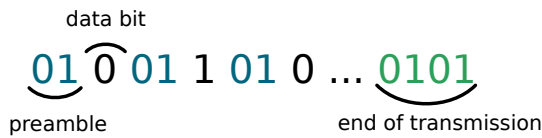}
\caption{The transmission protocol: 2 bits preamble before each actual data bit and 4 bits for end of transmission.}
\label{fig:protocol}
\end{figure}

Our protocol transmits one byte of data per LED for identification, with the lowest channel capacity available. 
The transmitter sends two preamble bits before every symbol, and four preamble bits to signal the end of transmission, making the full transmission 28 bits long for (see Figure~\ref{fig:protocol} ).
As the transmission is only one byte long, using a short preamble actually yields similar results compared to using a single long preamble, and ensures recovery of every symbol even in degraded conditions, while making the interference region easier to isolate. 

Regarding the modulation, Manchester coding is usually favored for its ability to code clock data into the signal. However, on/off keying uses half of the bandwidth. With this protocol, on/off keying permits us to transmit data with similar efficiency as if we used Manchester Coding with a single long preamble.

We make sure that consecutive LEDs transmit orthogonal preambles so that intersecting preambles cancel each other and create a region without modulation. 

Let $sp_i(t)$ be the preamble signal, $st_i(t)$ the on off keying signal, $\phi$ the irradiation angle, $\theta$ the incidence angle, $T$ the period of modulation, $i$ the LED number and ${x_i}$ is transmitted bit by $i^{th}$ Led ($x_i \in [0,1]$). We can describe $sp_i(t)$ as follows:
\begin{equation}
sp_i(t) = \left\{
    \begin{array}{ll}
       \dfrac{1+(-1)^i}{2} & t<T \\
        \dfrac{1-(-1)^i}{2} & T<t<2T \\
    \end{array}
\right\},\ st_i={x_i\ t<T},\ 
\end{equation}

The full transmission can then be represented as:
\begin{equation}
	P_i(t) = \left\{
   		\begin{array}{ll}
        	sp_i(t) & if\ preamble \\
            st_i(t) & if\ sending\ bit \\
       	\end{array}
\right\}
\label{eq:preamble}
\end{equation}

\subsection{Receiver}

The receiver exploits the rolling shutter effect to recover the encoded signal. In
this section, we call X axis the axis perpendicular to the rolling shutter direction, and Y
axis the axis parallel to the shutter direction. Our X axis is therefore always the axis
parallel to the black stripes generated by the LEDs. 

The reception process is composed of three phases : 

\begin{itemize}
\item Detecting the different light areas.
\item Finding the interference regions for each area.
\item Demodulate the signal and recover the message.
\end{itemize}

\subsubsection{Light source detection}

The light source detection process is similar to Luxapose\cite{kuo2014luxapose}.
We first pass the image through a Gaussian filter to eliminate the dark areas caused by
modulation. 
Then we find the contours of the white areas to extract the different light sources. 
As the camera's shutter speed is fast, the ambient light is filtered out, isolating the
transmitting light sources.
We then proceed to find the interference region for each light source.

\subsubsection{Finding interference region}

The CMOS camera of a smartphone captures the reflected light from a surface. To
recover the signal and detect the interferences, we focus on the relative light
intensity distribution on surface. 

\begin{figure}[t]
\centering
\includegraphics[width=0.25\textwidth]{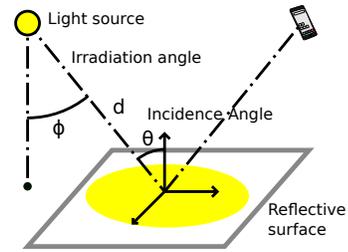}
\caption{The Lambertian radiance pattern}
\label{fig:lambertian}
\vspace{-0.8em}
\end{figure}

We use the Lambertian radiance pattern as the channel model (Figure~\ref{fig:lambertian}). Let $A$ be the Area of the receiver, $d$ the distance of the light source to the surface, $\phi$ the irradiation angle, $\theta$ the incidence angle and $m$ the Lambertian coefficient (constant, depends on cut off angle of LED, m=1 for $60^o$). We consider the equation for a single LED $i$ The radiance $H$ can be expressed as~\cite{kahn1997wireless}:
\begin{equation}
H_i(0) = A\dfrac{m+1}{2\pi}\dfrac{cos(\phi)^mcos(\theta)}{d_i^2}
\label{eq:lambert}
\end{equation}

To simplify the equation we consider $C_1=A\dfrac{m+1}{2\pi}$ and assume the surface and the light sources are perpendicular to each other. Then $\theta = \phi$ The channel equation becomes:
\begin{equation}
H_i(0)=\dfrac{C_1}{d_i^2}cos(\phi)^{m+1}
\end{equation}

And from~\ref{eq:preamble} and~\ref{eq:lambert} the light intensity is:
\begin{multline}
s_r(d,t)=\sum_i{H_i(0)P_i(t)+s_a(t)}\\
=\sum_i{\dfrac{P_i(t)C_1}{d_i^2}cos(\phi_i)^{m+1}+s_a(t)}
\end{multline}
$s_a(t)$ representing the ambient light.

With the rolling shutter mechanism, light is captured line by line, and only the spatial location is available in the X direction. However columns sample both spatial and temporal information.

\begin{figure}[t]
\centering
\includegraphics[width=0.45\textwidth]{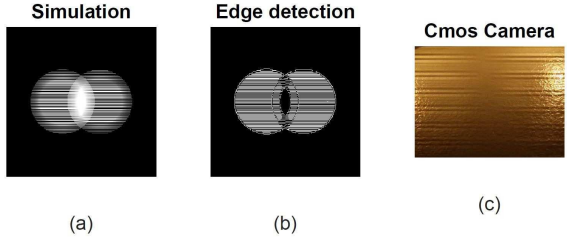}
\caption{a) Simulated LEDs with an interference region. b) edge detection algorithm applied to the simulated image, high frequency content at interference region is minimal. c) an image taken by a smart phone camera. }
\label{fig:Interference}
\vspace{-0.8em}
\end{figure}

We simulated two overlapping light regions with random signals. The results are presented on Figure~\ref{fig:Interference}. On Figure~\ref{fig:Interference}.a is represented the simulated LEDs, with the interference region in the center. Figure~\ref{fig:Interference}.b represents the same region after applying the edge detection algorithm. As we can see, the high frequency content of the interference region is negligible, leaving a distinguishable mark in the center. Finally, Figure~\ref{fig:Interference}.c shows the actual picture of 2 interfering LEDs, to validate the results of the simulation.

We detect interference region by calculating the average of the energy levels of high frequency content over the X axis.  

We start by computing the average energy levels over a window on the Y axis. Let $T_d$ be the amount of pixels required to transmit one bit of data. $T_d=\dfrac{T}{T_s}$, with T the modulation period and $T_s$ the sampling period. To ensure the presence of at least a single preamble in the window, its length has to be larger than $3T_d$.
The average of the window is then subtracted to the signal eliminate its DC content. We finally compute the remaining energy of the signal as follows:
\begin{equation}
E=\sum_{n=0}^{3T_d}{|X[n]|^2}
\end{equation}

\begin{figure}[t]
\centering
\includegraphics[width=0.45\textwidth]{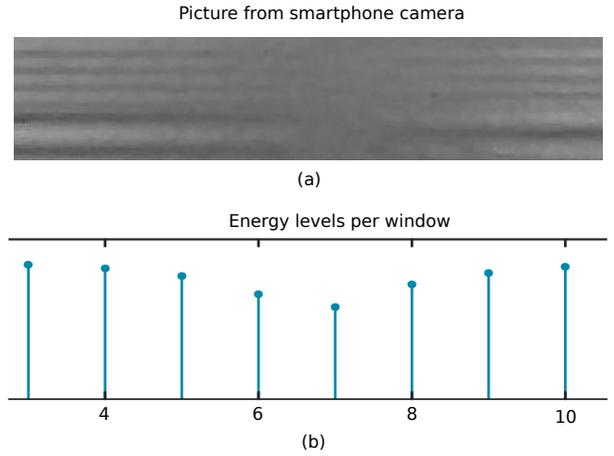}
\caption{a) interference region captured with a smartphone camera b) Energy levels of window slices through X axis. The energy levels drop around the interference zone}
\label{fig:region detection}
\vspace{-0.8em}
\end{figure}

We then repeat this operation over the X axis and compute the sliding average of the resulting energy levels.
In Figure~\ref{fig:region detection}, we
can observe how the energy levels change according to the regions: at the transmission region of LED $i$, the high frequency
content is higher, thus the energy levels are also higher. When getting closer to the interference region, the energy
levels start to drop, then rise again when approaching the transmission region of LED $i+1$. The local minimum points of the graph will give the center of the
interference regions, where SNR is lowest.

Once the regions of transmission are detected, the algorithm considers them as sub images, and decodes them separately. 
As our system transmits only one byte of information per LED, we decided to process the light sources linearly: the first region sends
the first part of the signal, the second region sends the second etc. 

Similarly, we solve the problem of orientation by always using an even number of LEDs.
As we use two orthogonal preambles, the signal starts with the LED showing the first preamble and ends with the last LED to use the other one.

\subsubsection{Demodulation}

The demodulation of the signal is done along the Y axis. First the DC content is filtered out and 2D image is converted to 1D image. 
As the lighting conditions may vary along Y axis, we made the choice to remove the DC component by subtracting the moving average to the signal instead of using the full signal mean. The window size of the moving average filter is chosen to be $3*T_n$ to include preamble into mean calculation. 
\begin{equation}
f(z) = \left\{
    \begin{array}{ll}
       0.5 & z>0 \\
        -0.5 & z<0\\
    \end{array}
\right\}
\end{equation}
where z is the current pixel value on gray-scale.

\begin{figure}[t]
\centering
\includegraphics[width=0.45\textwidth]{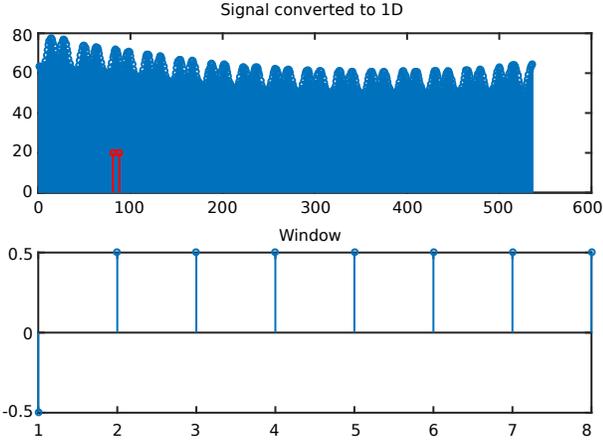}
\caption{On top the the signal array converted 1D is given. DC content is not filtered out yet, but modulation is visible. Second panel shows the processed signal. function.}
\label{fig:decode}
\vspace{-0.5em}
\end{figure}

As the channel noise is relatively low in visible light communication, the threshold function does not cause large jumps between sample points. The whole decoding process is shown on Figure~\ref{fig:decode}.

To recover the clock, we use a digital early late clock recovery algorithm. The frequent preamble signals satisfy the required high to low, low to high transitions for the algorithm to work.

%% file: evaluation.tex
\section{Evaluation}
\label{sec:eval}
In this section we evaluate the parameters that may affect the performance of our method, both analytically and experimentally.
To analyze how the distance between Light sources affects the interference regions we assume both sources are identical, located at equal height $h$ from the surface, and have equal cut off angle of 60 degrees. We assume $\phi,\theta$ are equal to each other for computation simplicity. The channel equation becomes:
    \begin{equation}
    cos(\phi)=cos(\theta)=\dfrac{h}{\sqrt{x^2+y^2+h^2}}
    \end{equation}
    \begin{equation}
	H_i(0) = \dfrac{C_1cos(\phi)cos(\theta)}{h^2+x^2+y^2} = \dfrac{C_1h^2}{(h^2+x^2+y^2)^2}
    \label{eq:channel}
    \end{equation}
    
From~\ref{eq:channel} we observe that the region characteristics depend on the height (distance to surface) $h$ and $d_{xy}=\sqrt{x^2+y^2}$, the position of the LED relatively to its reflection in the 2D plane. 

    To analyze the effect of $d_xy$ and $h$ we simulate the light intensity model with Matlab. In a first time, we consider $h$ as constant and vary $d_{xy}$ then we consider $d_{xy}$ as a constant and vary $h$. 

\subsection{Effect of distance between Light sources}
\label{sec:theory}

\begin{figure}[t]
\centering
\includegraphics[width=0.45\textwidth]{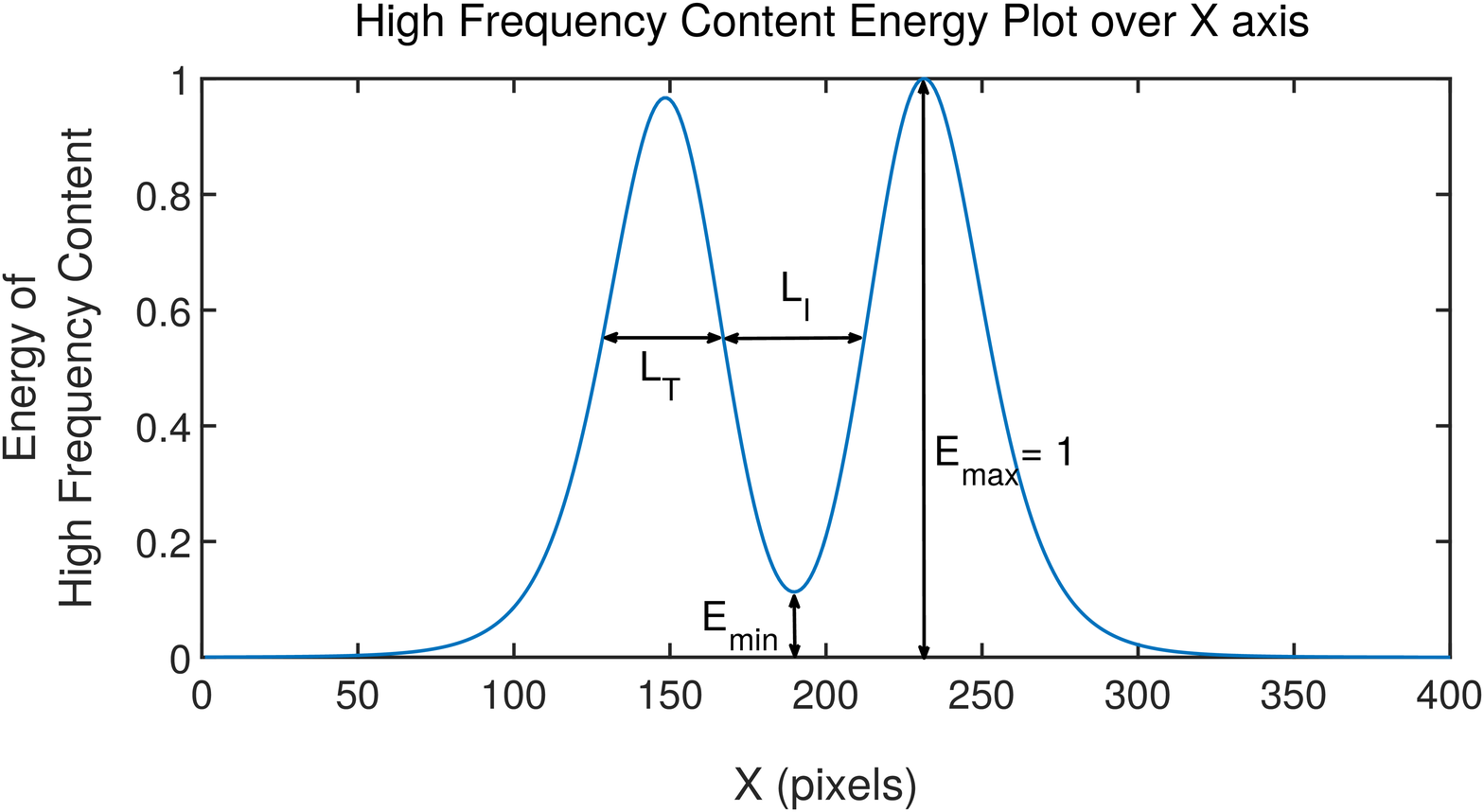}
\caption{Simulated energy plot $L_t$ corresponds to length of transmission region in pixels, $L_i$ corresponds to length of interference region in pixels.}
\label{fig:EnergySim}
\vspace{-0.5em}
\end{figure}

\begin{figure}[t]
\centering
\includegraphics[width=0.45\textwidth]{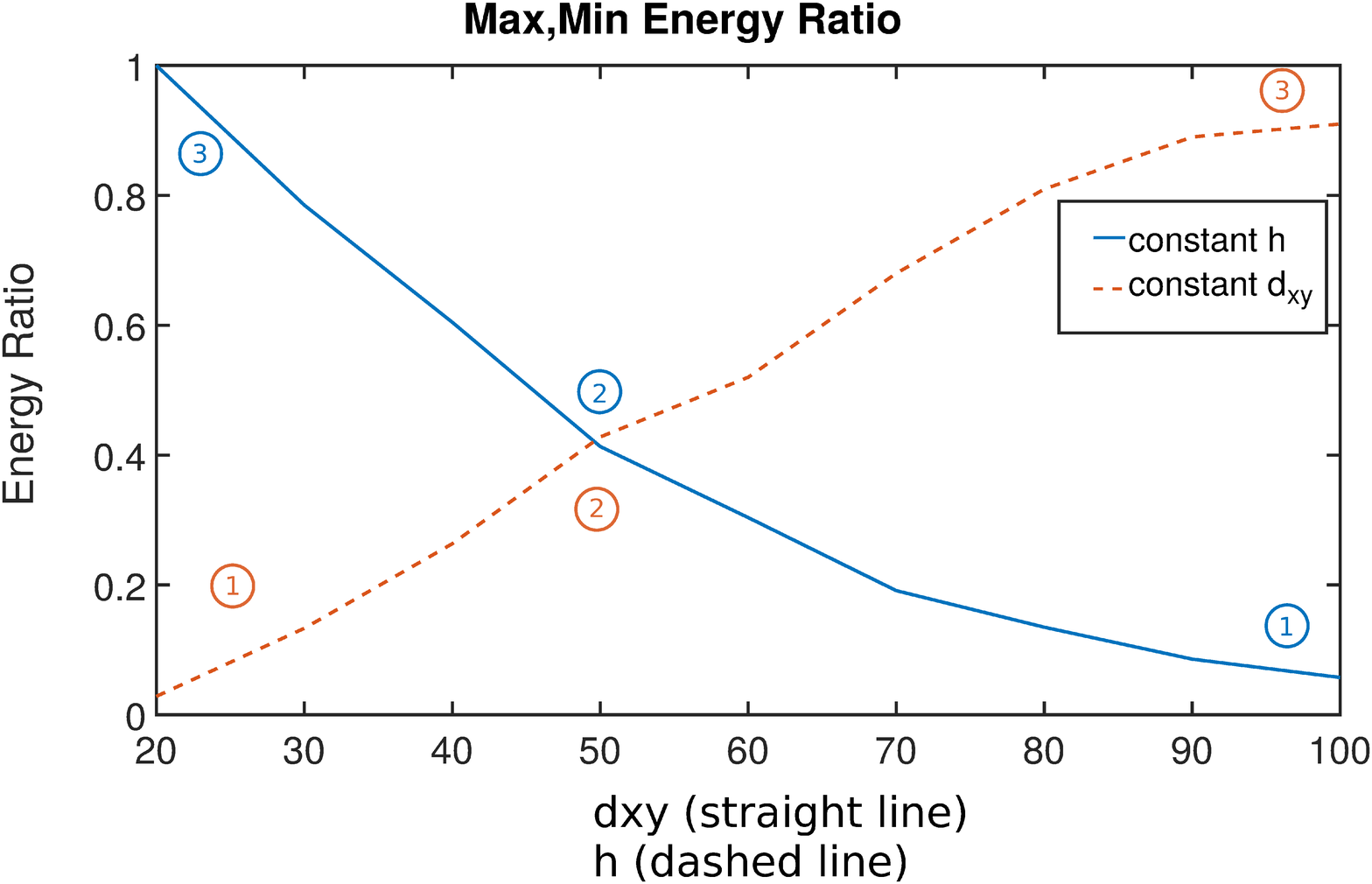}
\caption{Energy ratio of maximum and minimum points, which corresponds to transmission and interference regions. h = 50. As the ratio decreases it is easier to detect regions.}
\label{fig:EnergyRatio}
\vspace{-0.5em}
\end{figure}

In this section, we compare the normalized energy ratio between the transmission regions and the interference regions, relatively to $h$ and $d_{xy}$.
This energy ratio can be computed as $\dfrac{1}{E_{min}}$, $E_{min}$ being the local minimum energy level between two energy peaks (see Figure~\ref{fig:EnergySim}). The results of the simulations are displayed Figure~\ref{fig:EnergyRatio}.

\begin{figure}[t]
\centering
\includegraphics[width=0.45\textwidth]{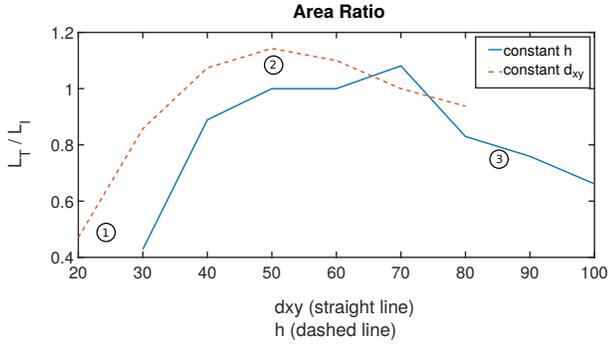}
\caption{Area ratio plot. }
\label{fig:AreaRatio}
\vspace{-0.5em}
\end{figure}

We also compare the size of the interference region relatively to the size of the transmission regions for various $h$ and$d_{xy}$. To do so, we first define a cut off energy level as $E_c = \dfrac{1+E_{min}}{2}$. Using this cut off energy we can define both the transmission and the interference regions as:    
    \begin{equation}
    \left\{
    \begin{array}{ll}
       transmission & E>E_c \\
       interference & E<E_c\\
    \end{array}
\right\}
    \end{equation}
When light sources are close to each other or when h gets larger, different light sources act as a single point source. Therefore, the transmission and interference regions cannot be separated. We display the results of this simulation Figure~\ref{fig:AreaRatio}.

Both Figures~\ref{fig:EnergyRatio} and~\ref{fig:AreaRatio} display the same phenomenon. The behavior of increasing h and $d_{xy}$ is similar for both energy and area ratios. This suggests that interference region can be characterized by the ratio between h and $d_xy$.

We identify three cases, that we represented on Figures~\ref{fig:EnergyRatio} and~\ref{fig:AreaRatio}:
\begin{enumerate}
	\item Light sources are very close to each other, with high $\dfrac{h}{d_{xy}}$ and act as a point source.
    \item Light sources are at a medium distance, a clear interference region is created by the cut off angle.
    \item Light sources are at far distance to each other with wider cut off angle and low $\dfrac{h}{d_{xy}}$ ratio. In this case there is a large interference region with low energy.
\end{enumerate}

    In the first case, the light intensities of the sources are approximately equal on the illuminated surface. This causes the majority of the illuminated region to appear as an interference region. Recovering the transmitted code in these condition becomes challenging.
    
    In the second case, the interference region borders are clearly visible due to the sudden change in energy. It is easy to detect both interference regions and the transmission. This case can be represented in real life as a LED array sending parallel data.
    
    In the third case, we observe a smooth transition of energy levels between the interference and transmission regions. However the area of the interference region becomes larger than the transmission region as change on x,y locations is not affecting the total distance as much as changes in height.
        
Cases two and three show that the light intensity of the interference region is strongly dependent on the position of the LEDs, and can even get lower than the transmission region. In such a scenario, a method based on energy of frequency contents would be more robust instead of detecting interference by light intensity.

Moreover, Figure~\ref{fig:EnergyRatio} shows that varying the height or the xy location has contradictory effects. If the LEDs get too close to each other, the interference region gets difficult to differentiate  from the transmission regions. On the other hand, putting the LEDs at a lower height enables optimal conditions for interference detection. 

\subsection{Experimental illustration}

 \begin{figure}[h!t]
  \centering
  \includegraphics[width=0.45\textwidth]{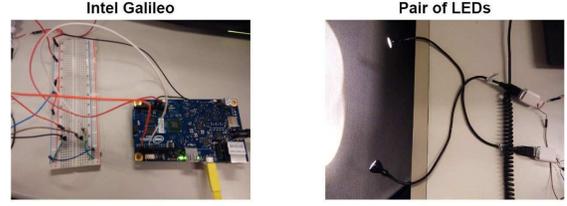}
  \caption{The experimental setup}
  \label{fig:setup}
  \vspace{-1em}
  \end{figure}

In this section, we aim to display the effect of various $h$ to $d_{xy}$ in real life conditions. To this purpose, we connected consumer market LEDs to an Intel Gallileo which handles the modulation and synchronization (see Figure~\ref{fig:setup}). The LEDs have 0.3\,W power and work with 5\,V from an USB outlet. The Intel Galileo modulates the LEDs at 8\,KHz. Due to the low power of the LEDs, we kept a low constant $h$=100\,mm. 

We ran the experiment in three different scenarios:
\begin{itemize}
	\item High $\frac{h}{d_{xy}}=5$, cutoff angle of 60\textdegree
    \item Medium $\frac{h}{d_{xy}}=2$, cutoff angle of 60\textdegree
    \item Low $\frac{h}{d_{xy}}=0.3$. In this case, the $h$ to $d_{xy}$ ratio is so high that we had to use a cutoff angle high enough to get both light sources to merge on the reflection surface.
\end{itemize}

  \begin{figure}[h!t]
  \centering
  \includegraphics[width=0.49\textwidth]{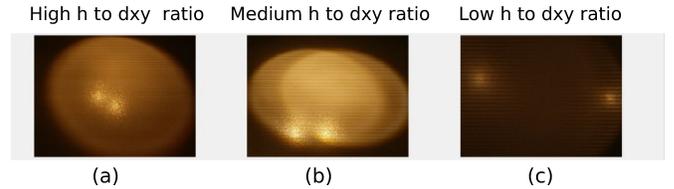}
  \caption{Interference Regions on image taken by a camera. (a) The LEDs act like single point source, the illuminated area is covered by the interference region. (b) Interference and transmission regions are clearly repeatable, (c) the interference region is at a very low energy area, due to limitations of LEDs and smart phone camera it is hard to capture a robust image from this area.}
  \label{fig:experiment}
  \vspace{-1em}
  \end{figure}

Figure~\ref{fig:experiment} displays the effect of high, medium and low $\frac{h}{d_{x}}$. In Figure~\ref{fig:experiment}.a, both LEDs are too close to each other compared to the distance to the surface (Zone A of Figure~\ref{fig:AreaRatio}), and the transmission region is hard to distinguish. On Figure~\ref{fig:experiment}.b, both interference and transmission regions are clearly distinguishable and the signal can be properly decoded.
Finally Figure~\ref{fig:experiment}.c displays the effect of a low  $\frac{h}{d_{xy}}$. As we are using low power LED and a short exposure time, the effect is harder to display on a picture. 
This shows what may happen when LEDs are too far away from each other, as it may happen for instance when illuminating an object from all sides in a display case. 
However, for most applications, we expect LEDs to be close to each other, resulting in the situations presented in Figures~\ref{fig:experiment}.a and~\ref{fig:experiment}.b.

This section confirms the theoretical results presented in Section~\ref{sec:theory} : the optimal ratio to detect the interference region is comprised between 0.5 and 2.

%% file: conclusion.tex
\section{Conclusion}
In this paper we have proposed a new method for finding spatial interference regions for VLC using off the shelf LEDs and the rolling shutter effect.  
Spatial interference detection can be used in AR applications for Ubiquitous networks and IoT, for instance as a discreet replacement for QR codes. 
Our method allows to increase the number of LEDs used to transmit a single message. By introducing such spacial parallelism, we are able to strikingly increase the bandwidth of the system.

	Our protocol is efficient for very shot transmissions such as IDs of devices, it contains too much overhead for a continuous flow.     
    In future works, we plan to design another protocol which would enable to exploit spatial parallelism for longer transmissions, and integrate it in high power lightbulbs in a similar way as LiFi~\cite{lifi}.  In order to achieve this goal, we should be able to take care of some problems we overlooked in our work such as the orientation of the interference region, as well as the situation where the interference region gets bigger than the transmission regions.
    
    \section{Acknowledgements}
This research has been supported, in part, by General Research Fund 26211515 from the Research Grants Council of Hong Kong, Innovation and Technology Fund ITS/369/14FP from the Hong Kong Innovation and Technology Commission.